\documentstyle[prd,aps,floats]{revtex}

\def\ds{\displaystyle}

\begin{document}

\draft
\input epsf
\twocolumn[\hsize\textwidth\columnwidth\hsize\csname
@twocolumnfalse\endcsname

\title{On the Moduli Problem and Baryogenesis in Gauge-mediated SUSY
Breaking Models}

\author{S. Kasuya, M. Kawasaki, and Fuminobu Takahashi}
\address{Research Center for the Early Universe, School of Science,
  University of Tokyo, Bunkyo-ku, Tokyo 113-0033, Japan}

\date{August 22, 2001}

\maketitle

\begin{abstract}
We investigate whether the Affleck-Dine mechanism can produce
sufficient baryon number of the universe in the gauge-mediated 
SUSY breaking models, while evading the cosmological moduli problem by
late-time entropy production. We find that the Q-ball formation
renders the scenario very difficult to work, irrespective of the
detail mechanism of the entropy production.
\end{abstract}

\pacs{PACS numbers: 98.80.Cq, 11.27.+d, 11.30.Fs
      \hspace{5cm} hep-ph/0108171}

\vskip2pc]

\setcounter{footnote}{1}
\renewcommand{\thefootnote}{\fnsymbol{footnote}}

\def\ds{\displaystyle}

\section{Introduction}
\label{sec:intro}

In superstring theories, there generally exist various dilaton and
modulus fields. These fields (we call them "moduli" in this paper) are
expected to acquire masses of the order of the gravitino mass
$m_{3/2}$ through some non-perturbative effects of supersymmetry
(SUSY) breaking~\cite{Carlos}. It is well known that the moduli cause
serious cosmological problem~\cite{Coughlan} because they have only
gravitationally suppressed interactions with other particles and hence
have long lifetimes. The moduli with mass $O(100)$ GeV decay at the Big
Bang Nucleosynthesis (BBN) epoch and spoil the success of the BBN by
destroying the synthesized light elements, while moduli with lighter
mass ($\lesssim 1$~GeV) may overclose the universe or emit
X($\gamma$)-rays  giving  too many contributions to the cosmic
background radiation~\cite{Kawasaki}. 

The mass of moduli (which is assumed to be the same as the gravitino
mass $m_{3/2}$) depends on models of SUSY breaking. In hidden sector
models~\cite{Nilles} the SUSY breaking in the hidden sector is
mediated by gravitation and SUSY particles (squarks, sleptons, etc )
in the observable sector as well as gravitino obtain the mass of weak
scale $\sim O(100)$~GeV. On the other hand, in gauge-mediated SUSY
breaking models~\cite{Giudice}, the gauge interactions mediate the
SUSY breaking effects. In gauge-mediated SUSY breaking, the gravitino
cannot acquires mass by the gauge interactions but only through
gravitation. Thus, the mass of gravitino is much lighter ($\lesssim
1$~GeV) than that in the hidden sector models.  The gauge mediation
models have attractive points that are absent in the hidden sector
models; they can avoid the flavor problem and the mass pattern of the
particles in the observable sector is predictable~\cite{Giudice}.
Therefore, in this paper we consider the moduli problem in the
gauge-mediated SUSY breaking models.

In order to avoid the moduli problem, we need some huge entropy
production process by which the moduli density is diluted.  So far the 
most successful mechanism for entropy production is ``thermal
inflation'' proposed by Lyth and Stewart~\cite{Lyth}.  The thermal
inflation model in the gauge-mediated SUSY breaking models was
intensively investigated in Refs.~\cite{Hashiba,Asaka1,Asaka2} where
it was shown that the thermal inflation can solve the moduli problem.
However, any process that dilutes the moduli also dilutes primordial
baryon asymmetry of the universe. Since the entropy production should
take place after the start of the moduli oscillation, the reheating
temperature is generally very low, which makes regeneration of the
baryon asymmetry almost impossible. Thus, we must produce sufficiently
large baryon asymmetry before the entropy production occurs. 

Although it is known that the GUT baryogenesis or the leptogenesis
could work for the mechanism for the baryogenesis, they could produce
the baryon numbers $\eta_B \sim 10^{-10}$ at most before the dilution.
The only promising candidate for mechanism of such efficient baryon
number generation is the Affleck-Dine (AD)
baryogenesis~\cite{Affleck}. In fact, it was shown in
Ref.~\cite{Asaka2} that both the present baryon asymmetry and small
moduli density can be explained by the thermal inflation and the
Affleck-Dine mechanism for the gauge-mediated SUSY breaking models.

However, it has been found that the dynamics of the Affleck-Dine
baryogenesis is complicated by the existence of
Q balls~\cite{Kusenko,Enqvist}. In the gauge-mediated SUSY breaking
models the potential for the Affleck-Dine field becomes flat at large
amplitudes. For such a flat potential the Q-ball formation is 
inevitable~\cite{Kasuya1,Kasuya2,Kasuya3}. In order to produce a large
baryon number the initial amplitude of the AD field should be large,
which also leads to the formation of Q balls with huge baryon($=$Q) 
number. Since large Q balls are stable, the baryon number may be
confined in the form of Q balls and there may exist very small baryon
asymmetry in the cosmic plasma, which means that the baryogenesis does
not work. 

Unstable Q balls can provide all the charges created before the charge 
trapping by the produced Q balls, but rather small amplitudes of the
AD field are necessary for the Q balls to decay into the ordinary
baryons, nucleons. Thus, sufficient baryon number is difficult to be
created from the beginning.

In this paper we study the cosmological moduli problem and
baryogenesis in the gauge-mediated SUSY breaking models taking into
account the Q-ball formation. It is found that the Q balls seriously
affect the Affleck-Dine baryogenesis and lower its efficiency. As
result we show that the AD baryogenesis hardly works in the presence
of the entropy production which is necessary to dilute the dangerous
moduli. 

\section{Moduli Problem}

Here we briefly discuss the moduli problem. The modulus field $\eta$
obtain a mass of the order of the gravitino mass $m_{3/2}$. During the 
primordial inflation the modulus field is expected to sit at some
minimum of the effective potential determined by the K\"ahler
potential and the Hubble parameter. In general, the minimum during the
inflation deviates from the true minimum of the moduli potential at
low energies and the difference of the two minimum is considered to be
of the order of the gravitational scale 
$M(=2.4\times 10^{18} \mbox{GeV})$. After the inflation, when the
Hubble parameter becomes comparable to the mass of the modulus, the
modulus field begins to roll down toward the true minimum and
oscillates. Then, the modulus density ($=$ oscillation energy) is
estimated as 
\begin{equation}
   \label{eq:moduli-dens}
   \frac{\rho_{mod}}{s} \simeq \frac{1}{8}T_{RH}
   \left(\frac{\eta_0}{M}\right)^2,
\end{equation}
where $s$ is the entropy density, $T_{RH}$ is the reheating
temperature and $\eta_0$ is the initial amplitude of the modulus
oscillation ($\eta_0 \sim M$). In deriving Eq.(\ref{eq:moduli-dens}), 
we have assumed that the modulus mass is equal to $m_{3/2}$ and the
reheating takes after the modulus field starts the oscillation. (When 
we estimate the baryon-to-entropy ratio later, the opposite case is
also considered. See below.) Since $T_{RH}$ should be higher than
about 10~MeV to keep the success of the BBN, the modulus-to-entropy
ratio is bounded from below,
\begin{equation}
   \frac{\rho_{mod}}{s}\gtrsim 1.25\times 10^{-3}{\rm GeV}.  
\end{equation}

The decay rate of the modulus is very small because it has only
gravitationally suppressed interaction. The lifetime is roughly
estimated as
\begin{equation}
    \tau_{\eta} \sim 10^{18}\sec 
    \left(\frac{m_{3/2}}{100{\rm MeV}}\right)^{-3}.
\end{equation}
Thus, for $m_{3/2} \lesssim 100$~MeV, the lifetime is longer than the
age of the universe and its present density much larger than the
critical density which is given by
\begin{equation}
    \frac{\rho_c}{s_0} = 3.6\times 10^{-9} h^2 {\rm GeV},
\end{equation}
where $h$ is the present Hubble parameter in units of
100km/sec/Mpc and $s_0(\simeq 2.8\times 10^3$~cm$^{-3})$ is the
present entropy density. The modulus with larger mass (100~MeV
$\lesssim m_{3/2} \lesssim $ 1~GeV) decays into photons whose flux
exceeds the observed background X(or $\gamma$)-rays. Therefore the
modulus is cosmological disaster and should be diluted by some entropy
production process. 

\begin{figure}[t!]
\centering
\hspace*{-7mm}
\leavevmode\epsfysize=7cm \epsfbox{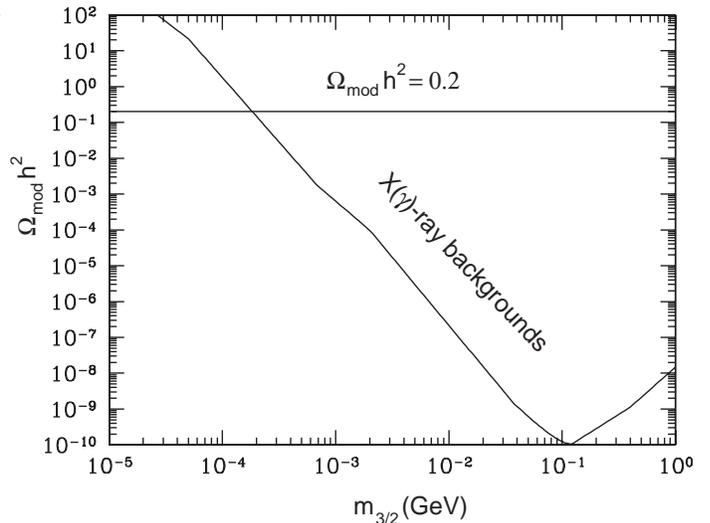}\\[2mm]
\caption[fig1]{\label{fig:obs} 
Observational upper limit of the density parameter of the modulus from 
observations. It is determined by the facts the modulus density should
not exceed the dark matter density for $m_{3/2} \lesssim$ 100 keV, and
the observed background X(or $\gamma$)-rays for 100 keV $\lesssim
m_{3/2} \lesssim$ 1 GeV.}
\end{figure}

Figure \ref{fig:obs} shows the observational upper limit of the
present density of the modulus. As mentioned above, the limit comes
from the facts that the modulus density should not exceed the dark
matter density for $m_{3/2} \lesssim$ 100 keV, and the observed
background X(or $\gamma$)-rays for 100 keV $\lesssim m_{3/2} \lesssim$
1 GeV. 

\section{Affleck-Dine mechanism and Q-ball formation}

In the Minimal Supersymmetric Standard Model (MSSM), there exist flat
directions, along which there are no classical potentials. Since flat
directions consist of squarks and/or sleptons, they carry baryon
and/or lepton numbers, and can be identified as the Affleck-Dine
field. These flat directions are lifted by SUSY breaking effects. In
the gauge-mediated SUSY breaking models, the potential of a flat
direction is parabolic at the origin, and almost flat beyond the
messenger scale~\cite{Gouvea,Kusenko,Kasuya3}: 
\begin{equation}
    V_{gauge} \sim \left\{ 
      \begin{array}{ll}
          m_{\phi}^2|\Phi|^2 & \quad (\Phi \ll M_S) \\
          \ds{M_F^4 \log \frac{|\Phi|^2}{M_S^2}}
          & \quad (\Phi \gg M_S) \\
      \end{array} \right.,
\end{equation}
where $M_{S}$ is the messenger mass scale.

Since the gravity always exists, flat directions are also lifted by
the gravity-mediated SUSY breaking effects \cite{EnqvistMcDonald98}:
\begin{equation}
    V_{grav} \simeq m_{3/2}^2 \left[ 1+K
      \log \left(\frac{|\Phi|^2}{M} \right)\right] |\Phi|^2,
\end{equation}
where $K$ is the numerical coefficient of the one-loop
corrections. This term can be dominant only at high energy scales
because of the small gravitino mass $\lesssim O(1\mbox{GeV})$. 

If the AD field directly couples with fields $\psi$ in the thermal
bath, it acquires a thermal mass term in the effective potential at
one-loop order: 
\begin{equation}
     V_T^{(1)} \sim f^2 T^2 |\Phi|^2,
\end{equation}
where $f$ is a Yukawa, or gauge coupling constant between the AD field
and the thermal particles which directly interact with it, and larger
than $10^{-5}$. Note that this effect is exponentially suppressed when
the effective mass of the thermal particle, $f |\Phi|$, is larger than
the temperature. This term can make the AD field oscillate earlier
than the case without this term, since the high reheating temperature
is possible in the absence of the gravitino problem. \footnote{
Since we consider the late-time entropy production for diluting the
modulus field, gravitino is also diluted. Thus, there is no
cosmological gravitino problem even if the reheating temperature is
high.}

In addition to this term, there is another thermal effect on the
potential, which appears at two-loop order, as pointed out in
Ref.~\cite{AnisimovDine}. This comes from the fact that the running of
the gauge coupling $g(T)$ is modified by integrating out heavy
particles which directly couples with the AD field. This contribution
is given by
\begin{equation}
     V_T^{(2)} \sim T^4 \log\frac{|\Phi|^2}{T^2}.
\end{equation}

The baryon number is usually created just after the AD field starts
coherent rotation in the potential, and its number density $n_B$ is
estimated as 
\begin{equation}
    n_B(t_{osc}) \simeq \varepsilon \omega \phi_{osc}^2,
\end{equation}
where $\varepsilon(\lesssim 1)$ is the ellipticity parameter, which
represents the strongness of the A-term, and $\omega$ and $\phi_{osc}$ 
are the angular velocity and the amplitude of the AD field at the
beginning of the oscillation (rotation) in its effective potential. 

Actually, however, the AD field feels spatial instabilities
during its coherent oscillation, and deforms into nontopological
solitons, Q balls \cite{Kusenko,Enqvist,Kasuya1}. In the case that the
zero-temperature potential $V_{gauge}$ dominates, the gauge-mediation
type Q balls are formed, whose properties are as follows \cite{Dvali}:
\begin{equation}
    \label{eq:mass}
    M_Q \sim M_F Q^{3/4}, \qquad R_Q \sim M_F^{-1} Q^{1/4},
\end{equation}
where $M_{Q}$ and $R_Q$ are the mass and the size of the Q ball,
respectively. If the mass per unit charge, $ M_F Q^{-1/4}$, is smaller
than the proton mass $\sim$ 1GeV, the Q ball is stable against the
decays into nucleons, which follows that Q balls with very large Q can
be stable. 

From numerical calculations~\cite{Kasuya1,Kasuya3}, Q balls absorb almost
all the baryon charges which the AD field obtains, and the typical
charge is estimated as \cite{Kasuya3}
\begin{equation}
    Q \simeq \beta \left(\frac{\phi_{osc}}{M_F}\right)^4,
\end{equation}
where $\beta \approx 6 \times 10^{-4}$. Consequently, the present
baryon asymmetry should be explained by the charges which come out of
the Q balls through the evaporation, diffusion, and decay of Q
balls. In the case that  $V_T^{(2)}$ dominates, one must replace
$M_F$ with $T_*$, where $T_*$ is the temperature at the Q-ball
formation. Notice that the shape of the Q ball reconfigures as the
temperature drops, releasing energy, but not charge. Finally, it will
become the zero-temperature configuration, as in Eq.(\ref{eq:mass}).

In the case of the unstable Q balls, they decay into nucleons and light
scalar particles. Since the temperature at the BBN time is very low
($\sim 1$ MeV), Q balls cannot decay into light scalars. The decay
rate is thus given by \cite{Coleman}
\begin{equation}
    \frac{dQ}{dt} \lesssim \frac{\omega^{3} A}{192 \pi^{2}},
\end{equation}
where $A$ is a surface area of the Q ball.\footnote{
For conservative estimation of the baryon density, we use the maximal
decay rate, i.e., $dt/dQ \simeq \omega^3A/(192\pi^2)$.}

In the case of the stable Q balls, the evaporation is the only way to
extract the baryon charges from Q balls. The total evaporated charge
from the Q ball is estimated as \cite{Laine,Banerjee,Kasuya3}, 
\begin{equation}
   \label{eq:chargeevp}
   \Delta Q \sim 10^{15}
   \left(\frac{m_{\phi}}{\mbox{TeV}}\right)^{-2/3}
   \left(\frac{M_F}{10^6\mbox{GeV}}\right)^{-1/3} Q^{1/12}.
\end{equation}
Hence the baryon number density is suppressed by the factor 
$\Delta Q/Q$, in comparison with the case of no stable Q-ball
production.

On the other hand, where $V_{grav}$ dominates the potential at larger
scales, the gravity-mediation type Q balls (`new' type) are produced
\cite{Kasuya2}, if $K$ is negative, while, if $K$ is positive, it is
not until the AD field enters $V_{gauge}$ dominant region that it
feels instabilities, and the gauge-mediation type Q balls are produced
(the delayed Q balls) \cite{Kasuya3}. Notice that the sign of $K$ is
in general indefinite in the gauge-mediated SUSY breaking models. We
will thus consider both cases later.

When the AD field starts to oscillate in the $V_{grav}$-dominant
region, where $H_{osc} \sim \omega \sim m_{3/2}$, the baryon number is
produced as $n_B \simeq \varepsilon m_{3/2}\phi_{osc}$. For the
negative $K$ case, the `new-type' Q balls are created, and its charge
is written as
\begin{equation}
    \label{eq:new-Q}
    Q \simeq \tilde{\beta} \left(\frac{\phi_{osc}}{m_{3/2}}\right)^2,
\end{equation}
where $\tilde{\beta} \simeq 6\times 10^{-3}$. This type of the Q ball
is also stable against the decay into nucleons, and the amount of the
baryons in the present universe is explained by the charge evaporation
from the Q balls. The charge evaporated from the Q ball is estimated
as \cite{Kasuya2} 
\begin{equation}
    \label{eq:new-evap}
    \Delta Q \sim 2.2 \times 10^{20} 
    \left(\frac{m_{3/2}}{100 \mbox{keV}}\right)^{-1/3}
    \left(\frac{m_{\phi}}{\mbox{TeV}}\right)^{-2/3}.
\end{equation}

Since the gauge-mediation type of the delayed Q ball are formed only
after the AD field enters the $V_{gauge}$-dominant region for the
positive $K$, the charge of the Q ball is given by
\begin{equation}
    \label{eq:delayq}
    Q \sim \beta \left(\frac{\phi_{eq}}{M_F}\right)^4
      \sim \beta \left(\frac{M_F}{m_{3/2}}\right)^4,
\end{equation}
where $\phi_{eq} \sim M_F^2/m_{3/2}$ is used. If $V_T^{(2)}$ dominates 
over the zero-temperature potential $V_{gauge}$, one must only replace
$M_F$ in the Eq.~(\ref{eq:delayq}) with $T_{eq}$. Here subscript `eq'
denotes the values when the gauge- (or thermal logarithmic) and
gravity-mediation potential are the same.

\section{Baryogenesis and the moduli problem}

As we mentioned in the Introduction, the late-time entropy production
necessary for the dilution of the moduli also dilutes the baryon
numbers created earlier very seriously, but the sufficient numbers
could remain, if the Q-ball production is not taken into account. We
will see that the Q-ball formation puts very serious restriction on
the efficiency of the AD baryogenesis, and makes it useless, whether
the produced Q balls are stable or not.

\subsection{Stable Q balls}
 
Since the baryon number is supplied only by the evaporation from
stable Q balls, it should be suppressed by the factor $\Delta Q/Q$,
compared with no Q-ball formation. We will show that this fact 
considerably reduces the power of the AD baryogenesis.


\subsubsection{Gauge-mediation type Q balls when the zero-temperature 
potential is dominated}

The AD field starts to oscillate when $H_{osc}\simeq
M_{F}^{2}/\phi_{osc}$. This is earlier than the beginning of the
moduli oscillation $H_{mod} \sim m_{3/2}$, since 
$M_F^4 \gtrsim m_{3/2}^2 \phi_{osc}^2$ in this case. There are two
situations when the moduli fields start the oscillation: before and
after the reheating. In the former case, the reheating temperature
should be lower than $T_{RH}^{(c)}$, which is defined as
\begin{eqnarray}
    \label{eq:reheating}
    T_{RH}^{(c)} & \equiv & 
    \left(\frac{90}{\pi^2 g(T_{RH})}\right)^{1/4} 
    \sqrt{m_{3/2} M}, 
    \nonumber \\ & \simeq &
     7.2 \times 10^6 \mbox{GeV} 
    \left(\frac{m_{3/2}}{100 \mbox{keV}}\right)^{1/2},
\end{eqnarray}
where $g(T_{RH})\sim 200$ counts the effective degrees of freedom of
the radiation. Since the ratio between $n_B$ and the energy of the
inflaton $\rho_{inf}$ stays constant until the reheating, we have
\begin{eqnarray}
     \frac{n_B}{\rho_{mod}}&=&\left. 
     \frac{n_B}{\rho_{inf}}\right|_{osc}\left.    
     \frac{\rho_{inf}}{\rho_{mod}}\right|_{H=m_{3/2}},
     \nonumber \\
     &\simeq&  \frac{n_B}{3H_{osc}^2 M^2} \times
     \frac{3m_{3/2}^2 M^2}{\frac{1}{2}m_{3/2}^2 M^2},
     \nonumber \\
     &=&\frac{2n_B}{H_{osc}^2 M^2}.
\end{eqnarray}

On the other hand, if the moduli start to oscillate after the
reheating, the baryon-to-moduli ratio at the beginning of the
moduli oscillation becomes larger by the factor $a_{mod}/a_{RH} \simeq 
T_{RH}/T_{mod}$, where subscript `mod' denotes the values at the
moduli oscillation time, since the universe is radiation-dominated
after the reheating until the moduli start to oscillate. We thus have
\begin{eqnarray}
     \frac{n_{B}}{\rho_{mod}} & \simeq &
     \left. \frac{n_{B}}{\rho_{inf}}\right|_{osc}
     \left. \frac{\rho_{rad}}{\rho_{mod}}\right|_{H=m_{3/2}}
     \times \frac{T_{RH}}{T_{mod}}, 
     \nonumber \\
     & \simeq & \frac{n_{B}}{3H_{osc}^{2}M^{2}} \
     \frac{3 m_{3/2}^{2}M^{2}}{\frac{1}{2}m_{3/2}^{2}M^{2}} 
     \times \frac{T_{RH}}{T_{RH}^{(c)}},
     \nonumber \\
     &=&  \frac{2n_{B}}{H_{osc}^{2}M^{2}} \times
     \frac{T_{RH}}{T_{RH}^{(c)}},
\end{eqnarray}
where $\rho_{rad}$ is the energy density of radiation. 

Therefore the ratio of the baryon number and the energy density of
the moduli can be written as
\begin{equation}
    \label{eq:ratio2}
    \frac{n_B}{\rho_{mod}}=\frac{2 n_B}{H_{osc}^2 M^2} \times r_H,
\end{equation}
where
\begin{eqnarray}
    r_H & = &\left\{
    \begin{array}{ll}
        1 & \quad \mbox{for} \quad T_{RH} < T_{RH}^{(c)} \\[2mm]
        \ds{\frac{T_{RH}}{T_{RH}^{(c)}}} & 
        \quad \mbox{for} \quad T_{RH} > T_{RH}^{(c)}
    \end{array}
    \right..
\end{eqnarray}

If the $V_{gauge}$ dominates over the thermal logarithmic potential 
$V_T^{(2)}$, $M_F \gtrsim T_{osc}$, so that the initial amplitude is
constrained as 
\begin{equation}
    \label{eq:t0pot}
    \phi_{osc} \gtrsim \left( \frac{T_{RH}}{M_F}\right)^2 M.
\end{equation}
$V_{gauge}$ also dominates over $V_{grav}$, which leads to the
condition
\begin{equation}
    \label{eq:v-gauge}
    \phi_{osc} \lesssim \frac{M_F^2}{m_{3/2}}.
\end{equation}
Combining these two equations, we have
\begin{equation}
    \label{eq:Trh1}
    T_{RH} \lesssim \frac{M_F^2}{\sqrt{m_{3/2}M}}.
\end{equation}

Since the scale of the SUSY breaking sector $\Lambda_{DSB}^{1/2}$ is
larger than $M_F$, the following condition should be hold:
\begin{equation}
    \label{eq:MF1}
    M_F \lesssim \Lambda_{DSB}^{1/2} \sim (m_{3/2}M)^{1/2},
\end{equation}
where the vanishing cosmological constant is assumed in the last
equality. From Eqs.(\ref{eq:Trh1}) and (\ref{eq:MF1}), the condition
on the reheating temperature becomes 
$T_{RH} \lesssim (m_{3/2}M)^{1/2}$. It corresponds to the case that
the modulus field starts its oscillation before the reheating, hence
$r_H=1$. 

Since $n_B \simeq \varepsilon \omega \phi_{osc}^2 \Delta Q/Q$,
and $H_{osc} \simeq \omega \simeq M_F^2/\phi_{osc}$, the
baryon-to-entropy ratio can be written as 
\begin{eqnarray}
    \label{eq:yb1}
    Y_B & = & \frac{n_B}{\rho_{mod}}\, 
    \frac{\rho_{mod}}{\rho_c}\, \frac{\rho_c}{s_0},
    \nonumber \\ & \simeq & 
    1.2 \times 10^{-30} \varepsilon
    \left(\frac{\Omega_{mod}h^2}{0.2}\right)
    \left(\frac{m_{\phi}}{\mbox{TeV}}\right)^{-2/3}
    \nonumber \\ & & \hspace{20mm} \times 
    \left(\frac{M_{F}}{10^6\mbox{GeV}}\right)^{4/3}
    \left(\frac{\phi_{osc}}{M}\right)^{-2/3},
    \nonumber \\  & \lesssim & 
    6.2 \times 10^{-27} \varepsilon 
    \left(\frac{\Omega_{mod}h^2}{0.2}\right)
    \left(\frac{m_{\phi}}{\mbox{TeV}}\right)^{-2/3},
\end{eqnarray}
where we put into the lower limit of $\phi_{osc}$ derived from the
stability condition: $M_Q/Q \lesssim 1$ GeV, which is expressed as
\begin{eqnarray}
    \label{eq:stable1}
    \frac{\phi_{osc}}{M} \gtrsim 2.6 \times 10^{-6}
    \left(\frac{M_{F}}{10^6\mbox{GeV}}\right)^2.
\end{eqnarray}

When we estimate the upper bound of $Y_B$, we can also put the
survival condition $\Delta Q \lesssim Q$, which reads as
\begin{equation}
    \label{eq:survive1}
    \frac{\phi_{osc}}{M} \gtrsim 3.2 \times 10^{-8}
    \left(\frac{m_{\phi}}{\mbox{TeV}}\right)^{-2/11}
    \left(\frac{M_{F}}{10^6\mbox{GeV}}\right)^{10/11},    
\end{equation}
for $M_F \lesssim M_F^*$, where $M_F^*$ is obtained by equating the
RHSs of Eqs.(\ref{eq:stable1}) and (\ref{eq:survive1}):
\begin{equation}
    M_F^* \simeq 1.8 \times 10^4 \mbox{GeV}
    \left(\frac{m_{\phi}}{\mbox{TeV}}\right)^{-1/6}.
\end{equation}
This leads to the upper bound on $Y_B$ as
\begin{eqnarray}
    Y_B & \lesssim & 1.2 \times 10^{-25} \varepsilon
    \left(\frac{\Omega_{mod}h^2}{0.2}\right)
    \nonumber \\ & & \hspace{15mm} \times
    \left(\frac{m_{\phi}}{\mbox{TeV}}\right)^{-6/11}
    \left(\frac{M_{F}}{10^6\mbox{GeV}}\right)^{8/11},
    \nonumber \\ & \lesssim & 6.2 \times 10^{-27} \varepsilon
    \left(\frac{\Omega_{mod}h^2}{0.2}\right)
    \left(\frac{m_{\phi}}{\mbox{TeV}}\right)^{-2/3},
\end{eqnarray}
where we put $M_F = M_F^*$ in the last line, since it make $Y_B$ be
maximum. Notice that this value is exactly the same as that
constrained by the stability condition, since the maximum $Y_B$ is
achieved at the boundary $M_F=M_F^*$ for the estimation derived using
the survival condition. 

In addition to the above case, complete evaporation of the Q ball can
be considered for appropriate parameters, and its condition is $\Delta
Q \gtrsim Q$, which is the opposite condition to
Eq.(\ref{eq:survive1}). The baryon-to-entropy ratio is 
\begin{eqnarray}
    Y_B & \simeq & 
    \frac{2\varepsilon\omega\phi_{osc}^2}{\omega^2 M^2}
    \Omega_{mod} \frac{\rho_c}{s_0},
    \nonumber \\ & \simeq & 3.5 \times 10^{-3} \varepsilon
    \left(\frac{\Omega_{mod}h^2}{0.2}\right)
    \left(\frac{M_{F}}{10^6\mbox{GeV}}\right)^{-2}
    \left(\frac{\phi_{osc}}{M}\right)^3,
    \nonumber \\ & \lesssim & 6.2 \times 10^{-27} \varepsilon
    \left(\frac{\Omega_{mod}h^2}{0.2}\right)
    \left(\frac{m_{\phi}}{\mbox{TeV}}\right)^{-2/3},
\end{eqnarray}
where $M_F=M_F^*$ is inserted in the last line. In either case, the
baryon-to-entropy ratio is too small to explain the present value of
the order $10^{-10}$.


\subsubsection{Gauge-mediation type Q balls when the thermal
logarithmic potential is dominated}

In this case, $V_{T}^{(2)}$ is dominant over $V_{gauge}$, and the AD
field starts to oscillate when $H_{osc} \sim T_{osc}^2/\phi_{osc}$. We
should also use the charge of the formed Q ball as $Q\simeq \beta
(\phi_{osc}/T_{osc})^4$. Therefore, the fraction of the evaporated
charge is written as
\begin{eqnarray}
    \frac{\Delta Q}{Q} & \sim & 10^{15} 
    \left(\frac{m_{\phi}}{\mbox{TeV}}\right)^{-2/3}
    \left(\frac{M_F}{10^6 \mbox{GeV}}\right)^{-1/3} 
    Q^{-11/12},
    \nonumber \\ & \sim & 8.9 \times 10^{17}
    \left(\frac{m_{\phi}}{\mbox{TeV}}\right)^{-2/3}
    \left(\frac{M_F}{10^6 \mbox{GeV}}\right)^{-1/3}
    \nonumber \\ & & \hspace{20mm} \times
    \left(\frac{T_{RH}}{M}\right)^{11/3}
    \left(\frac{\phi_{osc}}{M}\right)^{-11/2}.
\end{eqnarray}
Then the baryon-to-entropy ratio becomes
\begin{eqnarray}
    Y_B & \simeq & 5.3 \times 10^{-10} \varepsilon r_H
    \left(\frac{\Omega_{mod}h^2}{0.2}\right)
    \left(\frac{m_{\phi}}{\mbox{TeV}}\right)^{-2/3}
    \nonumber \\ & & \quad \times
    \left(\frac{M_F}{10^6\mbox{GeV}}\right)^{-1/3}
    \left(\frac{T_{RH}}{M}\right)^{5/3}
    \left(\frac{\phi_{osc}}{M}\right)^{-3/2}.
    \nonumber \\ & & 
\end{eqnarray}

Stability condition constrains the initial amplitude of the AD field
as
\begin{equation}
    \label{eq:stable2}
    \frac{\phi_{osc}}{M} \gtrsim 3.4 \times 10^4
    \left(\frac{M_{F}}{10^6\mbox{GeV}}\right)^{2/3}
    \left(\frac{T_{RH}}{M}\right)^{2/3}.
\end{equation}
This relation can be applied only to the situation when the following
condition holds:
\begin{equation}
    \frac{T_{RH}}{M} \lesssim 1.6\times 10^{-7} 
    \left(\frac{M_{F}}{10^6\mbox{GeV}}\right)^{-1},
\end{equation}
which comes from $\phi_{osc} \lesssim M$. On the other hand, there is
another constraint on $\phi_{osc}$, which comes from the survival
condition. It reads as
\begin{eqnarray}
    \label{eq:survive2}
    \frac{\phi_{osc}}{M} & \gtrsim & 1.8 \times 10^3
    \left(\frac{m_{\phi}}{\mbox{TeV}}\right)^{-4/33}
    \nonumber \\ & & \hspace{15mm} \times
    \left(\frac{M_{F}}{10^6\mbox{GeV}}\right)^{-2/33}
    \left(\frac{T_{RH}}{M}\right)^{2/3},
\end{eqnarray}
where, if this condition is applied, the RHS should be less than
unity, which is expressed as
\begin{equation}
    \label{eq:survive-T}
    \frac{T_{RH}}{M} \lesssim 1.3 \times 10^{-5} 
    \left(\frac{m_{\phi}}{\mbox{TeV}}\right)^{2/11}
    \left(\frac{M_{F}}{10^6\mbox{GeV}}\right)^{1/11}.
\end{equation}
Notice that the survival condition constrains more strictly on
$\phi_{osc}$ for $M_F \lesssim M_F^* \simeq 1.8\times 10^4
(m_{\phi}/\mbox{TeV})^{-1/6}$ GeV. 

It is easily seen that the largest upper limit on $Y_B$ comes from the 
survival condition, and we have
\begin{eqnarray}
    Y_B & \lesssim & 6.7 \times 10^{-15} \varepsilon r_{H}
    \left(\frac{\Omega_{mod}h^2}{0.2}\right)
    \left(\frac{m_{\phi}}{\mbox{TeV}}\right)^{-16/33}
    \nonumber \\ & & \quad \times
    \left(\frac{M_{F}}{10^6\mbox{GeV}}\right)^{-8/33}
    \left(\frac{T_{RH}}{M}\right)^{2/3}.
\end{eqnarray}

For the reheating temperature lower than $T_{RH}^{(c)}$, $r_H=1$, and
$T_{RH} \lesssim T_{RH}^{(c)}$ should be used for the upper bound on
$T_{RH}$, not Eq.(\ref{eq:survive-T}). Therefore, the upper limit on
the baryon-to-entropy ratio becomes
\begin{eqnarray}
    Y_B & \lesssim & 1.4 \times 10^{-22} \varepsilon
    \left(\frac{\Omega_{mod}h^2}{0.2}\right)
    \left(\frac{m_{\phi}}{\mbox{TeV}}\right)^{-16/33}
    \nonumber \\ & & \qquad \times
    \left(\frac{m_{3/2}}{100\mbox{keV}}\right)^{1/3}
    \left(\frac{M_{F}}{10^6\mbox{GeV}}\right)^{-8/33},
    \nonumber \\ & \lesssim & 7.4 \times 10^{-22} \varepsilon
    \left(\frac{\Omega_{mod}h^2}{0.2}\right)
    \left(\frac{m_{\phi}}{\mbox{TeV}}\right)^{-16/33}
    \nonumber \\ & & \hspace{25mm} \times
    \left(\frac{m_{3/2}}{100\mbox{keV}}\right)^{1/3},
\end{eqnarray}
where we put $M_F=1$ TeV, since it leads to the possible maximum
limit. Thus, it is too small to explain the present value. 

\begin{figure}[t!]
\centering
\hspace*{-7mm}
\leavevmode\epsfysize=7cm \epsfbox{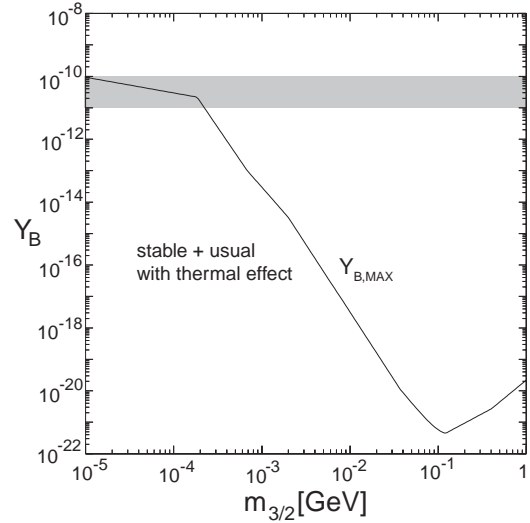}\\[2mm]
\caption[fig2]{\label{fig:stusu} 
Largest possible baryon-to-entropy ratio in the stable Q-ball
scenario in the thermal logarithmic potential.
}
\end{figure}

On the other hand, when the reheating temperature is higher than
$T_{RH}^{(c)}$, Eq.(\ref{eq:survive-T}) gives the upper bound on
$T_{RH}$, and the baryon-to-entropy ratio is estimated as
\begin{eqnarray}
    Y_B & \lesssim & 1.6 \times 10^{-11} \varepsilon
    \left(\frac{\Omega_{mod}h^2}{0.2}\right)
    \left(\frac{m_{\phi}}{\mbox{TeV}}\right)^{-2/11}
    \nonumber \\ & & \qquad \times
    \left(\frac{m_{3/2}}{100\mbox{keV}}\right)^{-1/2}
    \left(\frac{M_{F}}{10^6\mbox{GeV}}\right)^{-1/11},
\end{eqnarray}
where we use Eq.(\ref{eq:survive-T}) and
\begin{equation}
    \label{eq:r-h}
    r_H \simeq 3.4 \times 10^{11} 
    \left(\frac{m_{3/2}}{100\mbox{keV}}\right)^{-1/2}
    \left(\frac{T_{RH}}{M}\right).
\end{equation}
Thus, the largest possible upper limit is achieved at $M_F=1$ TeV as
\begin{eqnarray}
    Y_B & \lesssim & 3.0 \times 10^{-11} \varepsilon
    \left(\frac{\Omega_{mod}h^2}{0.2}\right)
    \left(\frac{m_{\phi}}{\mbox{TeV}}\right)^{-2/11}
    \nonumber \\ & & \hspace{30mm} \times
    \left(\frac{m_{3/2}}{100\mbox{keV}}\right)^{-1/2}. 
\end{eqnarray}
We plot the baryon-to-entropy
ratio in the function of $m_{3/2}$ in Fig.~\ref{fig:stusu}. We can thus
marginally explain the present value ($Y_B \gtrsim 10^{-11}$) for
$m_{3/2} \lesssim 100$ keV, but the reheating temperature is $\sim 2.0
\times 10^{13}$ GeV, too high for natural inflation models to provide.  

When the Q balls evaporate completely, the baryon-to-entropy ratio
can be obtained as 
\begin{eqnarray}
    Y_B & \lesssim & 1.4 \times 10^{-22} \varepsilon 
    \left(\frac{\Omega_{mod}h^2}{0.2}\right)
    \left(\frac{m_{\phi}}{\mbox{TeV}}\right)^{-16/33}
    \nonumber \\ & & \qquad \times
    \left(\frac{m_{3/2}}{100\mbox{keV}}\right)^{1/3}
    \left(\frac{M_{F}}{10^6\mbox{GeV}}\right)^{-8/33},
\end{eqnarray}
for the case that the modulus field starts its oscillation before
reheating ($T_{RH} \lesssim T_{RH}^{(c)}$), which is too small to
explain the present value. For the opposite case 
($T_{RH} \gtrsim T_{RH}^{(c)}$), we have 
\begin{eqnarray}
     Y_B & \lesssim & 1.6 \times 10^{-11} \varepsilon 
    \left(\frac{\Omega_{mod}h^2}{0.2}\right)
    \left(\frac{m_{\phi}}{\mbox{TeV}}\right)^{-2/11}
    \nonumber \\ & & \qquad \times
    \left(\frac{m_{3/2}}{100\mbox{keV}}\right)^{-1/2}
    \left(\frac{M_{F}}{10^6\mbox{GeV}}\right)^{-1/11},
\end{eqnarray}
which is marginally enough amount to explain the present
baryon-to-entropy ratio, although the reheating temperature should be
very high in order for this value to be achieved. Notice that it is
the same as for the case that the Q balls survive from the
evaporation, since $Y_B$ is maximized at $\Delta Q \sim Q$ in both
cases.


\subsubsection{Delayed Q balls when the zero-temperature 
potential is dominated}

Since both the AD field and modulus field start to oscillate when
$H_{osc}\sim m_{3/2}$, the ratio of the baryon number and the
energy density of the modulus stays constant to the present.
The baryon number is given by 
$n_B \simeq \varepsilon m_{3/2} \phi_{osc}^2 \Delta Q/Q$,
and the baryon-to-entropy ratio becomes
\begin{eqnarray}
    \label{eq:etab}
     Y_B & \sim & 2.8 \times 10^{-24} \varepsilon
     \left(\frac{\Omega_{mod}h^2}{0.2}\right)
     \left(\frac{m_{\phi}}{\mbox{TeV}}\right)^{-2/3}
     \nonumber \\ & & \hspace{5mm} \times
     \left(\frac{m_{3/2}}{100\mbox{keV}}\right)^{8/3}
     \left(\frac{M_F}{10^6\mbox{GeV}}\right)^{-4}
     \left(\frac{\phi_{osc}}{M}\right)^2,
\end{eqnarray}
where Eqs.(\ref{eq:chargeevp}) and (\ref{eq:delayq}) are used. The
gravitino mass is restricted by the stability condition, which can be
expressed as $m_{3/2} \lesssim 0.16$ GeV.

Since survival condition sets the upper limit on $Y_B$, we will
consider only this condition. It reads as
\begin{equation}
    \label{eq:mf3}
    \frac{M_F}{10^6\mbox{GeV}} \gtrsim 2.1\times 10^{-5}
    \left(\frac{m_{\phi}}{\mbox{TeV}}\right)^{-1/6}
    \left(\frac{m_{3/2}}{100\mbox{keV}}\right)^{11/12}.
\end{equation}
This can be applied if 
\begin{equation}
    \label{m32-3}
    m_{3/2} \gtrsim 6.8
    \left(\frac{m_{\phi}}{\mbox{TeV}}\right)^{2/11} \mbox{MeV}.
\end{equation}
Otherwise, we must use $M_F \gtrsim 1$ TeV, when we estimate the
upper bound on $Y_B$. Thus, we have
\begin{equation}
    Y_B \lesssim  1.1 \times 10^{-12} \varepsilon
    \left(\frac{\Omega_{mod}h^2}{10^{-6}}\right)
    \left(\frac{m_{3/2}}{6.8 \mbox{MeV}}\right)^{-1}
    \left(\frac{\phi_{osc}}{M}\right)^2,
\end{equation}
for $m_{3/2} \gtrsim 6.8 (m_{\phi}/\mbox{TeV})^{2/11}$ MeV, where
Eqs.(\ref{eq:mf3}) and (\ref{m32-3}) are used. We also take
$\Omega_{mod}h^2 \sim 10^{-6}$ (See Fig.\ref{fig:obs}). Notice that
this limit is same as the case for the complete evaporation of the Q
ball. On the other hand, when the gravitino mass is smaller than 
$6.8 (m_{\phi}/\mbox{TeV})^{2/11}$ MeV, 
\begin{eqnarray}
    Y_B & \lesssim & 2.8 \times 10^{-12} \varepsilon
    \left(\frac{\Omega_{mod}h^2}{0.2}\right)
    \left(\frac{m_{\phi}}{\mbox{TeV}}\right)^{-2/3}
    \nonumber \\ & & \quad \times
    \left(\frac{m_{3/2}}{100\mbox{keV}}\right)^{8/3}
    \left(\frac{M_F}{\mbox{TeV}}\right)^{-4}
    \left(\frac{\phi_{osc}}{M}\right)^2.
\end{eqnarray}

\begin{figure}[t!]
\centering
\hspace*{-7mm}
\leavevmode\epsfysize=7cm \epsfbox{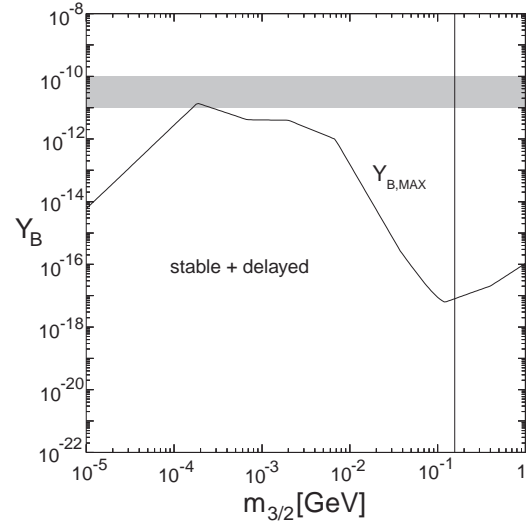}\\[2mm]
\caption[fig3]{\label{fig:delayed} 
Largest possible baryon-to-entropy ratio in the stable delayed Q-ball 
scenario. Notice that it can be applied in the zero-temperature
logarithmic potential $V_{gauge}$ for $m_{3/2} \lesssim 0.16$ GeV,
while all ranges of $m_{3/2}$ can be applied in the thermal logarithmic 
potential $V_T^{(2)}$.
}
\end{figure}

Figure \ref{fig:delayed} shows the maximum value of the
baryon-to-entropy ratio. As can be seen, this scenario is marginally
successful ($Y_B \sim 10^{-11}$) only for $m_{3/2} \sim 200$
keV. Notice that, the maximal value of $Y_B$ is the same as that in
the thermal logarithmic potential, to be considered in the next
subsection, for $m_{3/2} \lesssim 0.16$ GeV. 

\subsubsection{Delayed Q balls when the thermal
logarithmic potential is dominated}

We consider the case that $V_T^{(2)}$ is dominant over
$V_{gauge}$. This is the case if $T_{eq} \gtrsim M_F$, which is
satisfied when
\begin{equation}
    \label{eq:grav}
    \frac{\phi_{osc}}{M} \lesssim 
    \left( \frac{T_{RH}}{M_F}\right)^2.
\end{equation}
This condition can be applied when $T_{RH} \lesssim M_F$. Otherwise,
$\phi_{osc} \lesssim M$ should be used. $\phi_{osc}$ must be larger than 
$\phi_{eq}$ for the delayed Q-ball formation. It leads to
\begin{equation}
    \label{eq:thermal}
    \frac{\phi_{osc}}{M} \gtrsim 1.6\times 10^{11}
    \left(\frac{m_{3/2}}{100\mbox{keV}}\right)^{-1/2}
    \left(\frac{T_{RH}}{M}\right).
\end{equation}
This condition holds only if $T_{RH} \lesssim (m_{3/2}M)^{1/2}$. 
Combining Eqs.(\ref{eq:grav}) and (\ref{eq:thermal}), we have
\begin{equation}
    \frac{T_{RH}}{M} \gtrsim 2.6 \times 10^{-14}
    \left(\frac{m_{3/2}}{100\mbox{keV}}\right)^{-1/2}
    \left(\frac{M_F}{10^6\mbox{GeV}}\right)^2.
\end{equation}

Since the charge of the Q ball is written as
\begin{equation}
      Q=\beta \left(\frac{\phi_{eq}}{T_{eq}}\right)^4 \simeq
      \beta \left(\frac{T_{eq}}{m_{3/2}}\right)^4,
\end{equation}
the stability condition, given by $\omega \simeq M_F Q^{-1/4} \lesssim 
1$ GeV, is expressed as
\begin{eqnarray}
    \frac{\phi_{osc}}{M} & \lesssim & 1.4\times 10^{31}
    \left(\frac{m_{3/2}}{100\mbox{keV}}\right)^{-2}
    \left(\frac{M_F}{10^6\mbox{GeV}}\right)^{-2}
    \left(\frac{T_{RH}}{M}\right)^2,
    \nonumber \\
\end{eqnarray}
where this condition is effective only when the RHS is less than
unity, which leads to a constraint on the reheating temperature as 
\begin{equation}
    \frac{T_{RH}}{M} \lesssim 2.6 \times 10^{-16}
    \left(\frac{m_{3/2}}{100\mbox{keV}}\right)
    \left(\frac{M_F}{10^6\mbox{GeV}}\right).
\end{equation}
Otherwise, stability condition only implies that 
$\phi_{osc} \lesssim M$. In addition, the survival condition holds when
\begin{eqnarray}
    \frac{\phi_{osc}}{M} & \lesssim & 9.6 \times 10^{34}
    \left(\frac{m_{\phi}}{\mbox{TeV}}\right)^{4/11}
    \left(\frac{m_{3/2}}{100\mbox{keV}}\right)^{-2}
    \nonumber \\ & & \hspace{15mm} \times
    \left(\frac{M_F}{10^6\mbox{GeV}}\right)^{2/11}
    \left(\frac{T_{RH}}{M}\right)^2,
\end{eqnarray}
where this condition can be applied if
\begin{eqnarray}
    \frac{T_{RH}}{M} & \lesssim & 3.2 \times 10^{-18}
    \left(\frac{m_{\phi}}{\mbox{TeV}}\right)^{-2/11}
    \nonumber \\ & & \hspace{5mm} \times
    \left(\frac{m_{3/2}}{100\mbox{keV}}\right)
    \left(\frac{M_F}{10^6\mbox{GeV}}\right)^{-1/11}.
\end{eqnarray}
Otherwise, survival condition only implies that 
$\phi_{osc} \lesssim M$.

We must find the largest possible value of the baryon-to-entropy
ratio, 
\begin{eqnarray}
     Y_B & \simeq & 1.1 \times 10^{-69} \varepsilon
     \nonumber \\ & & \times
     \left(\frac{\Omega_{mod}h^2}{0.2}\right)
     \left(\frac{m_{\phi}}{\mbox{TeV}}\right)^{-2/3}
     \left(\frac{m_{3/2}}{100\mbox{keV}}\right)^{8/3}
     \nonumber \\ & & \times
     \left(\frac{M_F}{10^6\mbox{GeV}}\right)^{-1/3}
     \left(\frac{T_{RH}}{M}\right)^{-11/3}
     \left(\frac{\phi_{osc}}{M}\right)^{23/6},
\end{eqnarray}
in the parameter space constrained by the above five conditions:
thermal potential dominance, stability, survival, and delayed Q-ball
formation conditions, and $\phi_{osc} \lesssim M$. The maximum value
of $Y_B$ is achieved from the survival condition for $ m_{3/2} \gtrsim 
6.8 (m_{\phi}/\mbox{TeV})^{2/11}$ MeV, which is written as
\begin{equation}
    \label{eq:yb-st-th}
    Y_B \lesssim 1.1 \times 10^{-12} \varepsilon
    \left(\frac{\Omega_{mod}h^2}{10^{-6}}\right)
    \left(\frac{m_{3/2}}{6.8 \mbox{MeV}}\right)^{-1},
\end{equation}
and from the thermal potential dominance condition for $m_{3/2}
\lesssim 6.8 (m_{\phi}/\mbox{TeV})^{2/11}$ MeV, which can be expressed
as 
\begin{eqnarray}
    Y_B & \lesssim & 2.8 \times 10^{-12} \varepsilon
    \left(\frac{\Omega_{mod}h^2}{0.2}\right)
    \left(\frac{m_{3/2}}{100\mbox{keV}}\right)^{8/3}
    \left(\frac{m_{\phi}}{\mbox{TeV}}\right)^{-2/3},
    \nonumber \\ 
\end{eqnarray}
where we take $M_F=1$ TeV. These are plotted in
Fig.~\ref{fig:delayed}, and we can see that $Y_B$ is marginally enough
($Y_B \sim 10^{-11}$) only for $m_{3/2} \simeq 200$ keV. Notice that
the largest possible value of $Y_B$ is the same as
Eq.(\ref{eq:yb-st-th}) for the case that the charge of the Q ball
evaporates completely, since the largest value is achieved when 
$\Delta Q \sim Q$.


\subsubsection{New type Q balls}
From Eqs.(\ref{eq:new-Q}) and (\ref{eq:new-evap}), we have
\begin{equation}
    \frac{\Delta Q}{Q} \simeq 6.1 \times 10^{-23}
    \left(\frac{m_{\phi}}{\mbox{TeV}}\right)^{-2/3}
    \left(\frac{m_{3/2}}{100\mbox{keV}}\right)^{5/3}
    \left(\frac{\phi_{osc}}{M}\right)^{-2}.
\end{equation}
This leads to the baryon-to-entropy ratio as
\begin{eqnarray}
    Y_B & \lesssim & 8.8 \times 10^{-28} \varepsilon 
    \left(\frac{\Omega_{mod}h^2}{0.2}\right)
    \nonumber \\ & & \hspace{10mm} \times
    \left(\frac{m_{\phi}}{\mbox{TeV}}\right)^{-2/3}
    \left(\frac{m_{3/2}}{100\mbox{keV}}\right)^{2/3},
\end{eqnarray}
which is too small to explain the present value. Notice that this is
also true for the case of the complete evaporation, which condition
can be written as
\begin{equation}
    \frac{\phi_{osc}}{M} \lesssim 7.8 \times 10^{-12} 
    \left(\frac{m_{\phi}}{\mbox{TeV}}\right)^{-1/3}
    \left(\frac{m_{3/2}}{100\mbox{keV}}\right)^{5/6}.
\end{equation}


\subsection{Unstable Q balls}

In the case of the unstable Q balls which decay into nucleons, it may
destroy light elements synthesized at the BBN. Thus, a new constraint
which the Q ball should decay before the BBN ($\sim 1$ sec), must be
imposed.


\subsubsection{Gauge-mediation type Q balls when the zero-temperature 
potential is dominated} 

There are several condition to be imposed. The first is the condition
that the Q ball is unstable, given by 
\begin{equation}
    \label{eq:instability}
    \frac{\phi_{osc}}{M} \lesssim 2.6 \times 10^{-6} 
    \left(\frac{M_F}{10^6\mbox{GeV}}\right)^2. 
\end{equation}
Second, the decay of the Q ball must be completed until the BBN,
otherwise it would spoil the success of the BBN, so that the life 
time $\tau_Q$ should be
\begin{equation}
    \label{eq:lifetime}
     \tau_Q \equiv  \left(\frac{1}{Q} \frac{d Q}{dt}\right)^{-1}
     \simeq \frac{48 \pi}{M_F}Q^{5/4}  \lesssim   1 \mbox{sec},
\end{equation}
hence the following constraint:
\begin{equation}
    \label{eq:bbn}
     \frac{\phi_{osc}}{M} \lesssim 1.0 \times 10^{-6} 
     \left(\frac{M_F}{10^6 \mbox{GeV}}\right)^{6/5}.
\end{equation}
In addition, we have conditions Eqs.(\ref{eq:t0pot}) $-$ (\ref{eq:MF1}),
which leads to $r_H \simeq 1$. Notice that the
inequality~(\ref{eq:v-gauge}) is stronger than the
inequality~(\ref{eq:instability}), only for $m_{3/2}$ is larger than 
0.16 GeV. In either case, the largest possible value of $Y_B$ is
obtained by the conditions Eq.(\ref{eq:bbn}) and $M_F \lesssim
(m_{3/2}M)^{1/2}$, and will be written as
\begin{eqnarray}
    Y_B & \simeq & 3.5 \times 10^{-3} \varepsilon
    \left(\frac{\Omega_{mod}h^2}{0.2}\right)
     \left(\frac{M_F}{10^6\mbox{GeV}}\right)^{-2}
     \left(\frac{\phi_{osc}}{M}\right)^3
     \nonumber \\ & \lesssim & 2.8 \times 10^{-19} \varepsilon
    \left(\frac{\Omega_{mod}h^2}{0.2}\right)
    \left(\frac{m_{3/2}}{100\mbox{keV}}\right)^{4/5}.
\end{eqnarray}
This is thus too small to explain the present value $\sim 10^{-10}$.


\subsubsection{Gauge-mediation type Q balls when the thermal
logarithmic potential is dominated} 

Since the Q-ball charge is expressed as
\begin{equation}
    Q \simeq \beta \left(\frac{\phi_{osc}}{M}\right)^6 
    \left(\frac{T_{RH}}{M}\right)^{-4},
\end{equation}
the unstable condition, $M_F Q^{-1/4} \gtrsim 1$ GeV, is given by
\begin{equation}
    \frac{\phi_{osc}}{M} \lesssim 3.4\times10^4
    \left(\frac{M_F}{10^6 \mbox{GeV}}\right)^{2/3}
    \left(\frac{T_{RH}}{M}\right)^{2/3},
\end{equation}
while the lifetime condition that the Q ball decays before the BBN
time ($\sim 1$ sec), is written as
\begin{equation}
    \label{eq:lifetime2}
     \frac{\phi_{osc}}{M} \lesssim 1.9 \times10^4
     \left(\frac{M_F}{10^6 \mbox{GeV}}\right)^{2/15}
     \left(\frac{T_{RH}}{M}\right)^{2/3}.
\end{equation}
As will be seen, $Y_B$ becomes larger for larger $M_F$, so that the
lifetime condition determines the upper limit on $Y_B$. 

Let us first consider the case $r_H\simeq 1$, which sets the upper
bound on the reheating temperature. In general, the SUSY breaking
scenario sets the upper bound on $M_F$, such as $\lesssim
(m_{3/2}M)^{1/2}$. In this case, the RHS of Eq.(\ref{eq:lifetime2}) is 
less than unity, and the lifetime condition can be directly applied
for estimating the baryon-to-entropy ratio. Therefore, it will be
\begin{eqnarray}
    Y_B & \simeq & 6.0 \times 10^{-28} \varepsilon r_H
    \left(\frac{\Omega_{mod}h^2}{0.2}\right)
    \left(\frac{T_{RH}}{M}\right)^{-2} 
    \left(\frac{\phi_{osc}}{M}\right)^4,
    \nonumber \\ & \lesssim & 5.6 \times 10^{-18} \varepsilon
    \left(\frac{\Omega_{mod}h^2}{0.2}\right)
    \left(\frac{m_{3/2}}{100\mbox{keV}}\right)^{3/5},
\end{eqnarray}
where we use $T_{RH} \lesssim T_{RH}^{(c)}$ and $M_F \lesssim
(m_{3/2}M)^{1/2}$ in the last line. It is thus much smaller than the
present value. 

\begin{figure}[t!]
\centering
\hspace*{-7mm}
\leavevmode\epsfysize=7cm \epsfbox{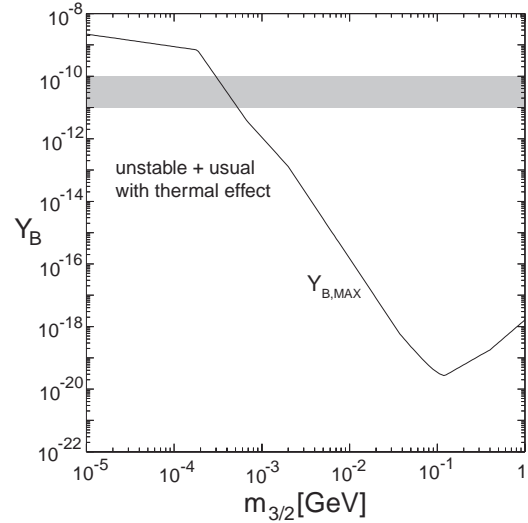}\\[2mm]
\caption[fig4]{\label{fig:unstusu} 
Largest possible baryon-to-entropy ratio in the unstable Q-ball
scenario in the thermal logarithmic potential.
}
\end{figure}

On the other hand, when the reheating temperature is higher than
$T_{RH}^{(c)}$, the lifetime condition again puts the upper limit on
$Y_B$. This condition can be applied if 
\begin{equation}
    \frac{T_{RH}}{M} \lesssim 4.1 \times 10^{-7}
    \left(\frac{M_F}{10^6 \mbox{GeV}}\right)^{-1/5}.
\end{equation}
Taking this constraint into account, we have the baryon-to-entropy
ratio as
\begin{equation}
    \label{eq:y-b-fig4}
    Y_B \lesssim 8.4 \times 10^{-10} \varepsilon 
    \left(\frac{\Omega_{mod}h^2}{0.2}\right)
    \left(\frac{m_{3/2}}{100\mbox{keV}}\right)^{-2/5},
\end{equation}
where $M_F \lesssim (m_{3/2}M)^{1/2}$ is again used. We plot the
largest possible value of $Y_B$ [Eq.(\ref{eq:y-b-fig4})] in the
function of $m_{3/2}$ in Fig.~\ref{fig:unstusu}. As can be seen, we
can  explain the present value for $m_{3/2} \lesssim 500$ keV. 
However, the reheating temperature should be as high as $5.6\times
10^{11}$ GeV, which may be rather too high for the actual inflation
models.

\subsubsection{Delayed Q balls}
When the zero-temperature potential, $V_{gauge}$, dominates over the
thermal logarithmic one, $V_T^{(2)}$, the unstable condition can be
expressed as $m_{3/2} \gtrsim 0.16$ GeV. On the other hand, when
$V_T^{(2)} \gtrsim V_{gauge}$, the same condition can be given by
\begin{equation}
    \frac{\phi_{osc}}{M} \gtrsim 1.4 \times 10^{31}
    \left(\frac{m_{3/2}}{100 \mbox{keV}}\right)^{-2}
    \left(\frac{M_F}{10^6 \mbox{GeV}}\right)^{-2}
    \left(\frac{T_{RH}}{M}\right)^2.
\end{equation}
In addition, the thermal potential dominance, $M_F \lesssim T_{osc}$,
is rewritten as
\begin{equation}
    \frac{\phi_{osc}}{M} \lesssim \left(\frac{T_{RH}}{M_F}\right)^2. 
\end{equation}
Combining these two, we obtain the constraint on the gravitino mass as
$m_{3/2} \gtrsim 0.16$ GeV. This is exactly the same as the former
case that the zero-temperature potential is dominant. Therefore, in
either case, we can estimate the baryon-to-entropy ratio as
\begin{equation}
    Y_B \lesssim 4.5 \times 10^{-17} \varepsilon
    \left(\frac{\Omega_{mod}h^2}{10^{-9}}\right)
    \left(\frac{m_{3/2}}{0.16\mbox{GeV}}\right)^{-1}
    \left(\frac{\phi_{osc}}{M}\right)^2,
\end{equation}
where $\Omega_{mod}h^2 \lesssim 10^{-9}$ for $m_{3/2} \gtrsim 0.16$
GeV (See Fig.~\ref{fig:obs}). This is too small to explain the present 
value, even if $\phi_{osc} \sim M$.


\subsection{Early oscillation due to the thermal mass term}

Since there is no cosmological gravitino problem because of the
late-time entropy production, the reheating temperature can be as
large as $\sim 10^{16}$ GeV. We take this value, because the COBE data 
implies that $V_{inf}^{1/4}/\epsilon_{sr}^{1/4} \simeq 6.7\times
10^{16}$ GeV \cite{LyRi}, where $\epsilon_{sr}$ is the slow-roll
parameter, and the instantaneous reheating is assumed for conservative
discussion. Therefore, the thermal mass term $V_T^{(1)} \sim f^2 T^2
\phi^2$ can extend towards as large as the Planck scale if the
coupling constant is not so large ($f \lesssim 4.1 \times
10^{-3}$). Here we consider the case that the thermal mass term causes
the early oscillation of the AD field. If the particles coupled to the
AD field are in the thermal bath, their mass should be less than the
temperature, so that  
\begin{equation}
    \label{eq:tm1}
    f \phi_{osc} \lesssim T_{osc},
\end{equation}
where $T_{osc}$ is the temperature at the beginning of the 
oscillation of the AD field. The oscillation starts when
$H_{osc} \sim f T_{osc}$, which reads as
\begin{equation}
    \omega \sim H_{osc} \sim \left(f^2 T_{RH} M^{1/2} \right)^{2/3}.
\end{equation}
Substituting this into Eq.~(\ref{eq:tm1}), we have
\begin{equation}
    \label{eq:tm3}
    \frac{\phi_{osc}}{M} \lesssim f^{-2/3}
    \left(\frac{T_{RH}}{M}\right)^{2/3}.
\end{equation}
This constraint can be used if its RHS is less than unity:
\begin{equation}
    \label{eq:tm5}
    T_{RH} \lesssim T_{RH}^* 
    \equiv 10^{-5} \left(\frac{f}{10^{-5}}\right) M.
\end{equation}
Otherwise, $\phi_{osc} \lesssim M$ and $T_{RH} \lesssim 10^{16}$ GeV
should be the only constraints. 

\begin{figure}[t!]
\centering
\hspace*{-7mm}
\leavevmode\epsfysize=7cm \epsfbox{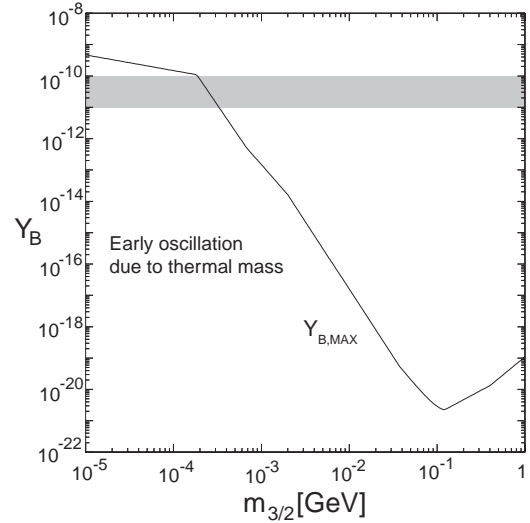}\\[2mm]
\caption[fig5]{\label{fig:thermal} 
Largest possible baryon-to-entropy ratio in the thermal mass term
dominance. We take $T_{RH} = 10^{16}$ GeV. 
}
\end{figure}

The baryon-to-entropy ratio is given by
\begin{eqnarray}
    \label{eq:tm4}
    Y_B & \simeq & \frac{2 \omega \phi_{osc}^2}{\omega^{2}M^{2}} r_H
    \Omega_{mod} \frac{\rho_c}{s_0}, 
    \nonumber \\ & \sim & 6.0 \times 10^{-28} 
    \varepsilon r_H f^{-4/3}
    \left(\frac{ \Omega_{mod}h^{2}}{0.2}\right)
    \nonumber \\ & & \hspace{20mm} \times
    \left(\frac{T_{RH}}{M}\right)^{-2/3}
    \left(\frac{\phi_{osc}}{M}\right)^2, 
    \nonumber \\ & \lesssim & 6.0 \times 10^{-28}
    \varepsilon r_H f^{-8/3}
    \left(\frac{ \Omega_{mod}h^{2}}{0.2}\right)
    \left(\frac{T_{RH}}{M}\right)^{2/3},
\end{eqnarray}
where we used Eq.~(\ref{eq:tm3}) in the last line. When the modulus 
field starts oscillation before the reheating ($r_H = 1$),
\begin{eqnarray}
    Y_B & \sim & 1.3 \times 10^ {-14} \varepsilon
    \left(\frac{f}{10^{-5}}\right)^{-8/3}
    \left(\frac{ \Omega_{mod}h^{2}}{0.2}\right)
    \left(\frac{T_{RH}}{M}\right)^{2/3},
    \nonumber \\ & \lesssim & 2.6 \times 10^{-22} \varepsilon
    \left(\frac{f}{10^{-5}}\right)^{\!\! -8/3} \!\!
    \left(\frac{ \Omega_{mod}h^{2}}{0.2}\right) \!
    \left(\frac{m_{3/2}}{100 \mbox{keV}}\right)^{1/3} \!, \nonumber \\ 
\end{eqnarray}
which is too small for the present baryon number. On the other hand,
when the modulus field starts its oscillation after the reheating, the 
baryon-to-entropy ratio becomes
\begin{eqnarray}
    Y_B & \lesssim & 2.0 \times 10^{-11} \varepsilon
    \left(\frac{f}{10^{-5}}\right)^{-1}
    \nonumber \\ & & \quad \times
    \left(\frac{ \Omega_{mod}h^{2}}{0.2}\right)
    \left(\frac{m_{3/2}}{100 \mbox{keV}}\right)^{-1/2}
    \left(\frac{T_{RH}}{T_{RH}^*}\right)^{5/3},
\end{eqnarray}
for $T_{RH} < T_{RH}^*$, where we use Eqs.(\ref{eq:r-h}) and
(\ref{eq:tm5}). If the reheating temperature is higher, we have 
\begin{eqnarray}
    \label{eq:yb-th}
    Y_B & \lesssim & 1.5 \times 10^{-10} \varepsilon
    \left(\frac{f}{10^{-5}}\right)^{-4/3}
    \left(\frac{ \Omega_{mod}h^{2}}{0.2}\right)
    \nonumber \\ & & \quad \times
    \left(\frac{m_{3/2}}{100 \mbox{keV}}\right)^{-1/2}
    \left(\frac{T_{RH}}{10^{16}\mbox{GeV}}\right)^{1/3}.
\end{eqnarray}
We show the upper limit on the baryon-to-entropy ratio in the function 
of $m_{3/2}$, Eq.(\ref{eq:yb-th}) in Fig.~\ref{fig:thermal}. We can
see that the present baryon number can be explained for 
$m={3/2} \lesssim 100$ keV. However, this reheating temperature is
unrealistically high. Moreover, since the Q-ball production will
diminish the efficiency of the baryon number creation, this value will
be much smaller, and cannot explain the present value.

\section{Conclusion}
We have investigate the possibility of the AD baryogenesis in the
gauge-mediated SUSY breaking scenario, while evading the cosmological
moduli problem by the late-time entropy production. In all the cases,
the Q-ball formation makes the efficiency of the baryon number
production considerably diminish. For the zero-temperature potential 
$V_{gauge}$-dominated case, whether the produced Q balls are stable or
not, the largest possible baryon-to-entropy ratio is too small to
explain the present value. This completely kills the successful
situations considered in Ref.~\cite{Asaka2}. 

We have also found that there are some marginally successful
situations when we take into account of the thermal effects on the
effective potential of the AD field. However, these successful
situations require very high reheating temperatures such as 
$10^{12} - 10^{16}$ GeV, which might be impossible to achieve in the
actual inflation models. Furthermore, Q balls must decay at the
maximal decay rate for these situations to be successful. However, it
is questionable whether such a fast decay process exists in the
gauge-mediated SUSY breaking models \cite{Yanagida}.

In the delayed Q-ball formation case, we have found that enough
baryon-to-entropy ratio can be created, even in the zero-temperature
potential $V_{gauge}$ is dominant over the thermal logarithmic
potential $V_T^{(2)}$. It might be the unique solution for the AD
baryogenesis with solving the cosmological moduli problem, although
the scale of $M_F$ is rather low.

In addition, successful situations above need the following
conditions: the large initial amplitude such as $\phi_{osc} \simeq M$, 
and $\varepsilon \simeq 1$. This is realized if the A-terms, which
make the AD field rotate in the effective potential, originate from
some K\"{a}hler potential with vanishing superpotential. Then, 
$\varepsilon \sim (\phi/M)^{\gamma} \sim 1$ for $\phi \sim M$, where
$\gamma > 2$. If the A-terms are determined by the nonrenormalizable
superpotential $W\sim \phi^n/M^{n-3}$, $\epsilon \sim 1$ can be
obtained in some parameter region, but the amplitude of the AD field
becomes much less than the Plank scale, i.e., $\phi_{osc} \ll M$,
which makes the baryon-to-entropy ratio much smaller than the present
value.  

If we consider the large late-time entropy production, the only
candidate we know is the thermal inflation models. In order for the
entropy production to be enough to dilute the dangerous moduli fields, 
the thermal inflation must last for long enough. This constrains the
scale of $M_F$ to be larger than $\sim 10^6$ GeV. If so, all the
successful scenario with not too high reheating temperatures found
here (i.e., the delayed Q-ball scenario) will not work. Therefore, we
can conclude that the AD baryogenesis is not compatible with the
late-time entropy production evading the cosmological moduli problem. 

Since the baryogenesis before the late-time entropy production does
not account for enough amount of the baryons in the present universe,
there should be some mechanism worked after the late-time entropy
production. If the reheating temperature after the late-time entropy
production is higher than the electroweak scale, the electroweak
baryogenesis might work. However, it is generally very difficult to
have strong first-order phase transition for the Higgs mass of 
$\simeq 114$ GeV or larger.

\section*{Acknowledgment}   
This work is supported by the Grant-in-Aid for Scientific Research
from the Ministry of Education, Science, Sports, and Culture of Japan,
Priority Area ``Supersymmetry and Unified Theory of Elementary
Particles'' (No.\ 707).


\begin{references}
\bibitem{Carlos}
  See, e.g. B. de Carlos, I.A. Casas, F. Queredo, and E. Roulet,
  Phys. Lett. \textbf{B318}, 447 (1993).

\bibitem{Coughlan}
  G.D. Coughlan, W. Fischler, E.W. Kolb, S.Raby, and G.G. Ross,
  Phys. Lett. \textbf{131B}, 59 (1983).

\bibitem{Kawasaki}
  M. Kawasaki and T. Yanagida,
  Phys. Lett. \textbf{B399}, 45 (1997).

\bibitem{Nilles}
  For a review, see, e.g. H.P. Nilles,
  Phys. Rep. \textbf{110}, 1 (1984).

\bibitem{Giudice}
  For a review, see, e.g. G.F. Giudice and R. Rattazzi,
  Phys. Rep. \textbf{322}, 419 (1999).

\bibitem{Lyth}
  D.H. Lyth and E.D. Stewart,
  Phys. Lett. \textbf{75}, 201 (1995);\\
  Phys. Rev. \textbf{D53}, 1784 (1996).

\bibitem{Hashiba}
  J. Hashiba, M. Kawasaki, and T. Yanagida,
  Phys. Rev. Lett. \textbf{79}, 4525 (1997).

\bibitem{Asaka1}
  T. Asaka, J. Hashiba, M. Kawasaki, and T. Yanagida,
  Phys. Rev. \textbf{D58}, 083509 (1998).

\bibitem{Asaka2}
  T. Asaka and M. Kawasaki,
  Phys. Rev. \textbf{D60}, 123509 (1999).

\bibitem{Affleck}
  I. Affleck and M. Dine,
  Nucl. Phys. \textbf{B249}, 361 (1985).

\bibitem{Kusenko}
  A. Kusenko and M. Shaposhnikov,
  Phys. Lett. \textbf{B418}, 46 (1998).

\bibitem{Enqvist}
  K. Enqvist and J. McDonald,
  Nucl. Phys. \textbf{B538}, 321 (1999).

\bibitem{Kasuya1}
  S. Kasuya and M. Kawasaki,
  Phys. Rev. \textbf{D61}, 041301 (2000).

\bibitem{Kasuya2}
  S. Kasuya and M. Kawasaki,
  Phys. Rev. Lett. \textbf{85}, 2677 (2000).

\bibitem{Kasuya3}
  S. Kasuya and M. Kawasaki, 
  to appear in Phys. Rev. \textbf{D}, 
  hep-ph/0106119.

\bibitem{Gouvea}
  A. de Gouv\^{e}a, T. Moroi, and H. Murayama,
  Phys. Rev. \textbf{D56}, 1281 (1997).

\bibitem{EnqvistMcDonald98}
  K. Enqvist and J. McDonald,
  Phys. Lett. \textbf{B425}, 309 (1998).

\bibitem{AnisimovDine}
  A. Anisimov and M. Dine,
  hep-th/0008058.

\bibitem{Dvali} G. Dvali, A. Kusenko, and M. Shaposhnikov,
  Phys. Lett. \textbf{B417}, 99 (1998).

\bibitem{Coleman}
  A. Cohen, S. Coleman, H. Georgei, and A. Manohar,
  Nucl. Phys. \textbf{B272}, 301 (1986).

\bibitem{Laine}
  M. Laine, and  M. Shaposhnikov,
  Nucl. Phys. \textbf{B532},376 (1998).

\bibitem{Banerjee}
  R. Banerjee and K. Jedamzik,
  Phys. Lett. \textbf{B484}, 278 (2000).

\bibitem{LyRi} D. Lyth and A. Riotto,
  Phys. Rep. \textbf{314}, 1 (1999).

\bibitem{Yanagida} T. Yanagida, {\it private communications.}

\end{references}
\end{document}